# ROBUST PID SLIDING MODE CONTROL FOR DC SERVO MOTOR SPEED CONTROL

**Ngoc Son Vu**[1*], **Van Cuong Pham**[1,2], **Phuc Anh Nguyen**[1],
**My Linh Dao Thi**[2], **Thanh Hai Vu**[2]

[1] *Hanoi University of Industry, Hanoi, Vietnam*
[2] *Thai Binh University, Thai Binh Province, Vietnam*
* vungocson.haui.271199@gmail.com

**Abstract:** This research proposes a Sliding Mode PID (SMC-PID) controller to improve the speed control performance of DC servo motors, which are widely used in industrial applications such as robotics and CNC. The objective of the proposed controller is to enhance the speed control performance of DC servo motors on the CE110 Servo Trainer. The proposed method integrates a traditional PID controller with a sliding mode control mechanism to effectively handle system uncertainties and disturbances. Experimental results show that the SMC-PID method provides significant improvements in accuracy and stability compared to traditional PID controllers, with metrics such as reduced overshoot, shorter settling time, and increased adaptability to system uncertainties. This research highlights the effectiveness of the SMC-PID controller, enhancing the performance of DC servo motor speed control.

## 1. INTRODUCTION

DC servo motors play a vital role in modern industrial systems, including robots, CNC machines, and precision drives. Their high accuracy and fast response make speed control a key research focus. However, practical applications often face challenges such as disturbances, parameter variations, and system uncertainties, which degrade control performance [1].

Among the various control methods currently applied, the PID control strategy has been maintained as a fundamental approach due to its simple structure, ease of implementation, and satisfactory performance in a wide range of applications. Despite its widespread use, certain limitations have been identified in conventional PID controllers, particularly when systems operate under uncertain conditions or are subjected to external disturbances [2,3]. To address these limitations, several advanced control strategies have been investigated and applied, including fuzzy logic control, sliding mode control (SMC), backstepping control, and other intelligent algorithms. Fuzzy logic control has been utilized to handle system nonlinearities and uncertainties effectively, particularly in robotic manipulator applications [4]. Sliding mode control has been recognized for its robustness against disturbances and model uncertainties in nonlinear dynamic systems [5]. Additionally, adaptive backstepping control techniques have been developed to enhance trajectory tracking performance in complex robotic systems under parameter variations and external disturbances [6].

In this study, a control scheme integrating the classical Proportional-Integral-Derivative (PID) controller with Sliding Mode Control (SMC), referred to as SMC-PID, has been proposed to enhance the speed regulation of DC servo motors implemented on the CE110 Servo Trainer platform. The approach has been formulated by embedding a sliding mode control law into the conventional PID framework in order to improve robustness and adaptability in the presence of system uncertainties and external disturbances [7-9]. The performance of the proposed controller has been experimentally validated and compared with that of the traditional PID controller. Improvements have been observed in terms of tracking accuracy, transient response, and overall system stability [7,9].

The research is structured as follows: ection 2 introduces the CE110 Servo Trainer and its MATLAB/Simulink model. Section 3 details the PID-SMC controller design. Section 4 presents experimental results comparing SMC-PID with conventional PID. Section 5 concludes and suggests future research directions.

## 2. SYSTEM DESCRIPTION

The experimental setup employs the CE110 Servo Control System by TecQuipment [14], comprising a DC motor, linear load, speed measurement unit, and an electrically controlled variable load, as depicted in Figure 1.

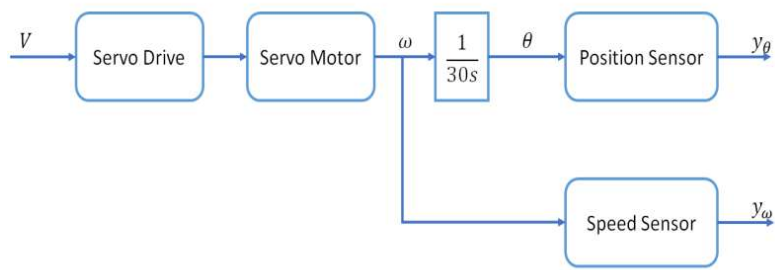

**Figure 1.** Block diagram of the feedback control system – CE110 Servo Trainer

In this system, $V$ denotes the input voltage applied to the motor, $\omega$ epresents the angular velocity of the motor shaft, and $\theta$ refers to the angular position of the shaft.

The output signals $y_\omega$ and $y_\theta$ correspond to feedback signals from the speed and position sensors, respectively [15].

The CE110 by TecQuipment, interfaced with a computer via the PCIe-6321 control board, ensures efficient communication and control. Motor speed is regulated by the input voltage, with an encoder measuring and transmitting transient speed data for analysis. The experimental setup includes a CE110 Servo Trainer, PCIe-6321 board, and a PC with MATLAB/Simulink.

The system, modeled in MATLAB/Simulink, operates with a 10ms sampling period for rapid data acquisition and feedback. An amplifier block converts voltage signals to motor speed, with a sensor gain calibrated at 1V per 200 RPM for precise control and monitoring.

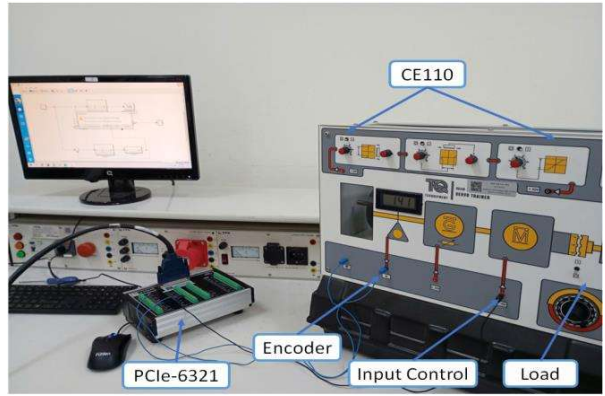

**Figure 2.** CE110 Servo Trainer system integrated with PCIe-6321 controller

### 3. PID-SMC CONTROLLER DESIGN

To enhance performance and robustness, a hybrid PID-SMC controller is proposed, combining the steady-state accuracy of PID with the robustness of SMC.

The control signal $u(t)$ consists of two main components:

$$u(t) = u_{PID}(t) + u_{SMC}(t) \qquad (1)$$

Where: $u_{PID}(t) = K_p e(t) + K_i \int e(t)dt + K_d \frac{de(t)}{dt}$ is the conventional PID control term. And $u_{SMC}(t) = -\eta \cdot sign(s(t))$ is the sliding mode control term designed to ensure robustness against uncertainties and external disturbances.

Here, $e(t) = \omega_{ref}(t) - \omega(t)$ is the speed error, and $\omega_{ref}(t)$ is the desired speed trajectory. The sliding surface is defined as:

$$s(t) = \lambda_1 e(t) + \lambda_2 \frac{de(t)}{dt} \qquad (2)$$

Where: $\lambda_1, \lambda_2$ are positive constants selected to shape the convergence dynamics of the error. To reduce chattering from the discontinuous sign function in practical scenarios, a saturation function or boundary layer approach is applied:

$$sign(s) \to sat\left(\frac{s}{\phi}\right) \quad (3)$$

To ensure system stability under the proposed PID-SMC controller, the Lyapunov candidate function is chosen as:

$$V(t) = \frac{1}{2}s^2(t) \quad (4)$$

Taking the time derivative:

$$\dot{V}(t) = s(t).\dot{s}(t) \quad (5)$$

Differentiate $s(t)$ with respect to time:

$$\dot{s}(t) = \lambda_1 \dot{e}(t) + \lambda_2 \ddot{e}(t) \quad (6)$$

The control input is designed as:

$$u(t) = u_{PID}(t) + u_{SMC}(t)$$

Assume the closed-loop system dynamics are of the form:

$$\ddot{e}(t) = f(e, \dot{e}) + b.u(t) \quad (7)$$

Substituting $u(t)$ into $\dot{s}(t)$, we have:

$$\dot{s}(t) = \lambda_1 \dot{e}(t) + \lambda_2 \left[f(e, \dot{e}) + b.\left(u_{PID}(t) - \eta.sat\left(\frac{s}{\phi}\right)\right)\right] \quad (8)$$

Group all bounded terms (including PID and nonlinear dynamics) into a function $\Delta(t)$ assumed to be bounded:

$$\Delta(t) = \lambda_1 \dot{e}(t) + \lambda_2 f(e, \dot{e}) + \lambda_2 b.u_{PID}(t) \quad (9)$$

Then: $\dot{V}(t) = s(t).\left[\Delta(t) - \lambda_2 b\eta.sat\left(\frac{s}{\phi}\right)\right] \quad (10)$

Now we choose $\eta$ large enough such that:

$$|\Delta(t)| \le \delta \to \eta > \frac{\delta}{\lambda_2 b} \quad (11)$$

Then: $\dot{V}(t) \le |s(t)||\Delta(t)| - \lambda_2 b\eta |s(t)| \left|sat\left(\frac{s}{\phi}\right)\right| \quad (12)$

Since $\left|sat\left(\frac{s}{\phi}\right)\right| \le 1$ then $\dot{V}(t) \le -\varepsilon |s(t)| < 0$ for some $\varepsilon > 0$

Thus, $\dot{V}(t) < 0$ proving that the closed-loop system is globally asymptotically stable in the Lyapunov and the tracking error converges to zero.

## 4. RESULTS

This section presents the experimental results of the PID-SMC controller applied to both the theoretical model [10] and practical experiments conducted on the CE110 Servo Trainer system [11]. The simulation and experimental setup diagrams were established and implemented in Matlab/Simulink to evaluate the performance of the PID-SMC controller.

The transfer function representing the system model:

$$G(s) = \frac{0,946}{1 + 0,4425s} e^{-0,0325s}$$

The parameters of the PID-SMC controller for the DC servo motor model, as calculated in Section 3, are:

$K_p = 25.1; K_i = 30.5; K_d = 0.293; \lambda_1 = 1; \lambda_2 = 0.05; \eta = 1; \emptyset = 0.05$

### 4.1. Simulation Results

The motor speed response was simulated under the following scenarios:

- The desired motor speed is 400 RPM

- The motor speed varies over time

The performance metrics of the PID-SMC control system were compared with those of a PID controller designed based on Kuhn synthesis theory and a randomly selected PID controller, as shown in Table 1.

*Bảng 1.* Bảng tham số PID

| Thông số | Kp | Ki | Kd |
|---|---|---|---|
| PID - Kuhn | 1.057 | 3.125 | 0.08016 |
| PID | 1 | 1 | 1 |

The results show that the PID-SMC controller achieves optimal speed response with zero overshoot, minimal settling time, and no steady-state error across all scenarios. It consistently outperforms PID-Kuhn and conventional PID controllers during acceleration, deceleration, and steady-state operation, demonstrating superior control performance and robustness.

### 4.2. Experimental Results

Experimental results show strong agreement, confirming the PID-SMC controller's effectiveness in both simulation and real-world applications. This consistency demonstrates its capability to track the desired speed with minimal error, validating its feasibility and robustness in practical use.

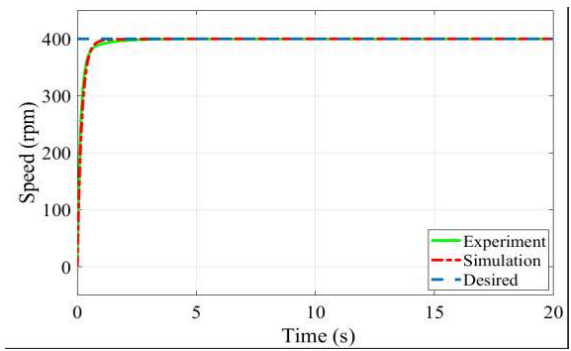

Figure 5. Motor speed response for desired speed of 400 RPM

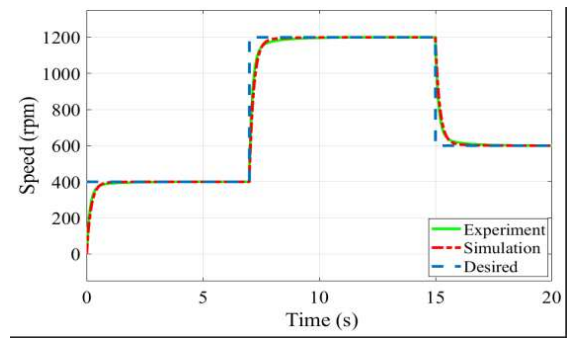

Figure 6. Experimental motor speed response with varying speed

## 5. CONCLUSION

This paper presented a robust SMC-PID controller for the speed control of DC servo motors on the CE110 Servo Trainer platform. By combining the traditional PID control strategy with the robustness of sliding mode control, the proposed approach effectively addressed the challenges of system uncertainties and external disturbances. Experimental results demonstrated that the SMC-PID controller significantly outperformed the conventional PID controller in terms of tracking accuracy, reduced overshoot, faster settling time, and improved stability. These results confirm the effectiveness and practicality of the SMC-PID controller in enhancing the speed regulation performance of DC servo systems. Future work will explore optimization techniques and adaptive mechanisms to further improve controller performance in more complex or highly nonlinear environments.

# BỘ ĐIỀU KHIỂN TRƯỢT - PID BỀN VỮNG CHO ĐIỀU KHIỂN TỐC ĐỘ ĐỘNG CƠ MỘT CHIỀU SERVO

**Ngoc Son Vu[1*], Van Cuong Pham[1,2], Phuc Anh Nguyen[1], My Linh Dao Thi[2], Thanh Hai Vu[2]**

[1] Hanoi University of Industry, Hanoi, Vietnam
[2] Thai Binh University, Thai Binh Province, Vietnam
* vungocson.haui.271199@gmail.com

**Tóm tắt:** Nghiên cứu này đề xuất bộ điều khiển PID trượt (SMC-PID) nhằm cải thiện hiệu suất điều khiển tốc độ động cơ servo DC, thành phần được ứng dụng rộng rãi trong các ứng dụng công nghiệp như Robot, CNC. Mục tiêu của bộ điều khiển đề xuất là nâng cao hiệu suất điều khiển tốc độ của động cơ servo DC trên bộ huấn luyện CE110. Phương pháp đề xuất tích hợp bộ điều khiển PID truyền thống với cơ chế điều khiển trượt để xử lý hiệu quả các thay đổi bất định và nhiễu loạn trong hệ thống. Kết quả thí nghiệm cho thấy phương pháp SMC-PID cung cấp cải thiện đáng kể về độ chính xác và ổn định so với PID truyền thống, với các chỉ số như giảm thiểu quá điều chỉnh, rút ngắn thời gian ổn định và tăng cường khả năng thích ứng với sự thay đổi bất định của hệ thống. Nghiên cứu này nhấn mạnh tính hiệu quả của bộ điều khiển SMC-PID, nâng cao hiệu suất cho việc điều khiển tốc độ động cơ servo DC.

**Từ khóa:** Điều khiển PID trượt, Kiểm soát mạnh mẽ, Động cơ servo DC, CE110 Servo Trainer.